\documentclass{PoS}
\usepackage{graphicx}
\usepackage{epstopdf}

\def\Det{{\rm Det}\,}

\def\Eq#1{Eq.~(\ref{#1})}
\renewcommand\Re{{\rm Re}\,}
\title{Why (staggered fermions)$^{1/4}$ fail at finite density}

\ShortTitle{Why (staggered fermions)$^{1/4}$ fail at finite density}

\author{\speaker{Benjamin Svetitsky} and
        Yigal Shamir\\
        School of Physics and Astronomy,
        Raymond and Beverly Sackler Faculty of Exact Sciences,
        Tel~Aviv University,
        69978 Tel~Aviv, Israel\\
        E-mail: \email{bqs@julian.tau.ac.il}, \email{shamir@post.tau.ac.il}}
\author{Maarten Golterman\\
        Department of Physics and Astronomy,
        San Francisco State University,
        San Francisco, CA~94132, USA\\
        E-mail: \email{maarten@stars.sfsu.edu}}

\abstract{Because the staggered fermion determinant is complex at nonzero $\mu$, taking its fourth root leads to phase ambiguities.  These unphysical effects cause the measure to become discontinuous; the problem becomes acute when $\Re\mu\gtrsim m_\pi/2$ (when $T>0$ this rough bound probably moves towards larger $\mu$).  We show how to overcome the problem, but only very close to the continuum limit.  This regime may be beyond reach with current resources.}

\FullConference{XXIV International Symposium on Lattice Field Theory\\
                July 23-28, 2006\\
                Tucson, Arizona, USA}

\begin{document}

All the recent numerical work on QCD at finite density has used staggered fermions, for well-known reasons of economy.  In order to get the right number of flavors one has to take the square root and the fourth root of the fermion determinant.  I will point out a fatal flaw in this procedure, one that will invalidate calculations performed at sufficiently large values of $\mu$ unless they are performed very close to the continuum limit.  The argument in a nutshell is as follows \cite{GSS}.

The determinant $\Delta[U]\equiv\Det[D(U)+m]$ may be written as a product of eigenvalues $\eta_i$ of the staggered Dirac operator $D(U)+m$.  The eigenvalues are complex, and they move as the gauge field $U$ is changed.  When $\Re\mu>0$ the determinant is complex.
A reweighting \cite{FK1,FK3} procedure needs the fourth root
$\Delta[U]^{1/4}$.  One has to define the phase of this root carefully.  For instance, if one takes the root of each $\eta_i$ in the cut plane, and defines $\Delta^{1/4}=\prod\eta_i^{1/4}$, then $\Delta^{1/4}$ will jump whenever any eigenvalue crosses the cut.

Are these jumps a bad thing?  I will show you an alternative prescription for $\Delta^{1/4}$ that has no jumps---and that is obviously correct.  So the jumps are indeed wrong, and a source of systematic error.
Our ``ideal'' prescription, on the other hand, only works if there is an almost unbroken taste symmetry in the low-energy spectrum, which is only true very near the continuum limit.  Far from this limit, taste symmetry is badly broken and there is no way to keep the jumps from returning, along with enormous systematic error.  We contend that current calculations are still too far from the limit, and that this will be the situation for the foreseeable future.

I will not touch on the recent progress in demonstrating the validity of the continuum limit for staggered fermions at $\mu=0$.  This is the subject of several talks at this conference \cite{CMY}. I recommend rooting around in the excellent summary by Steve Sharpe \cite{Sharpe}.

\section{Eigenvalues of the staggered Dirac operator}

\begin{figure}[thb]
\begin{center}
\includegraphics*[width=.4\columnwidth]{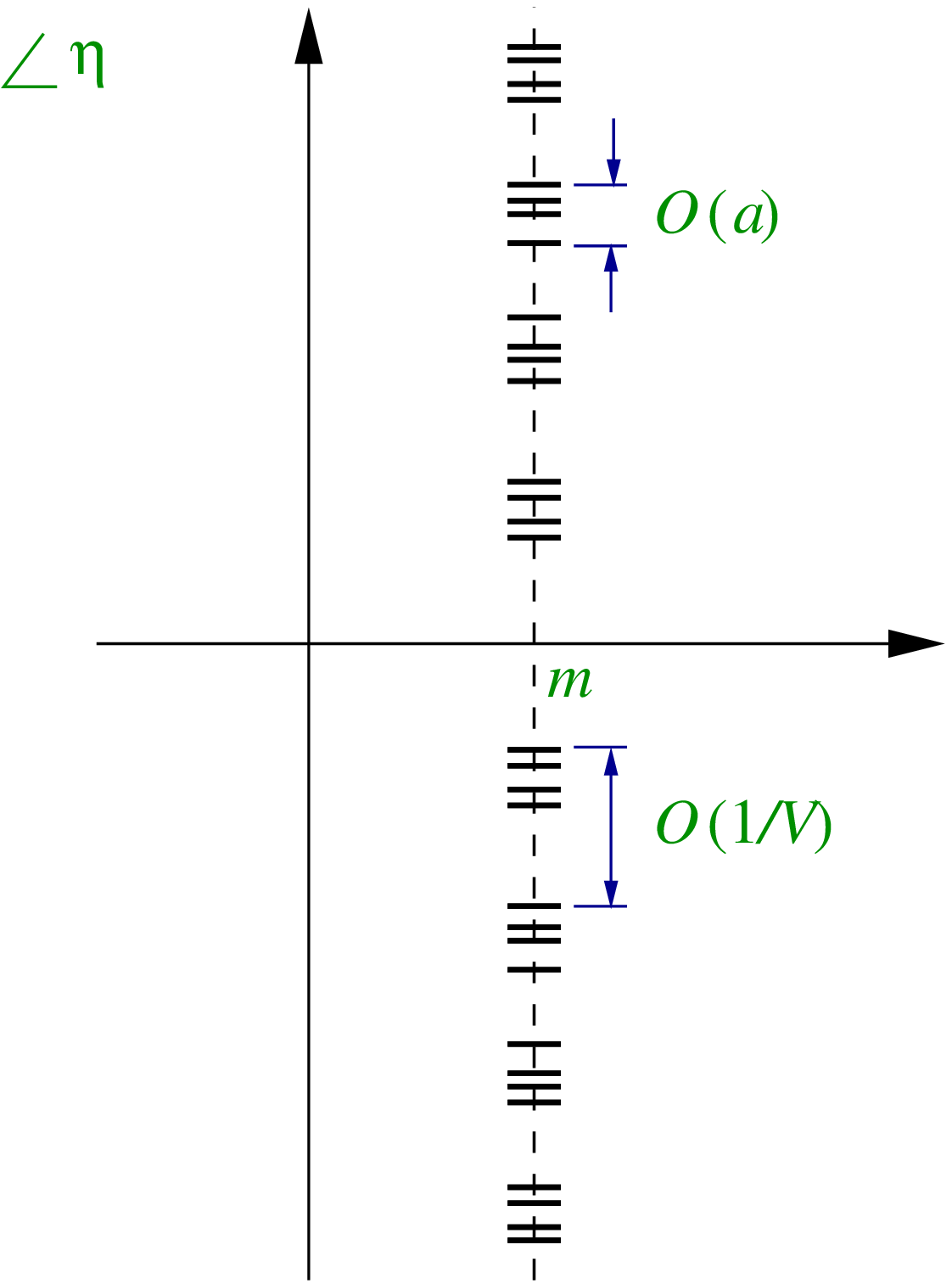}
\qquad\qquad
\includegraphics*[width=.4\columnwidth]{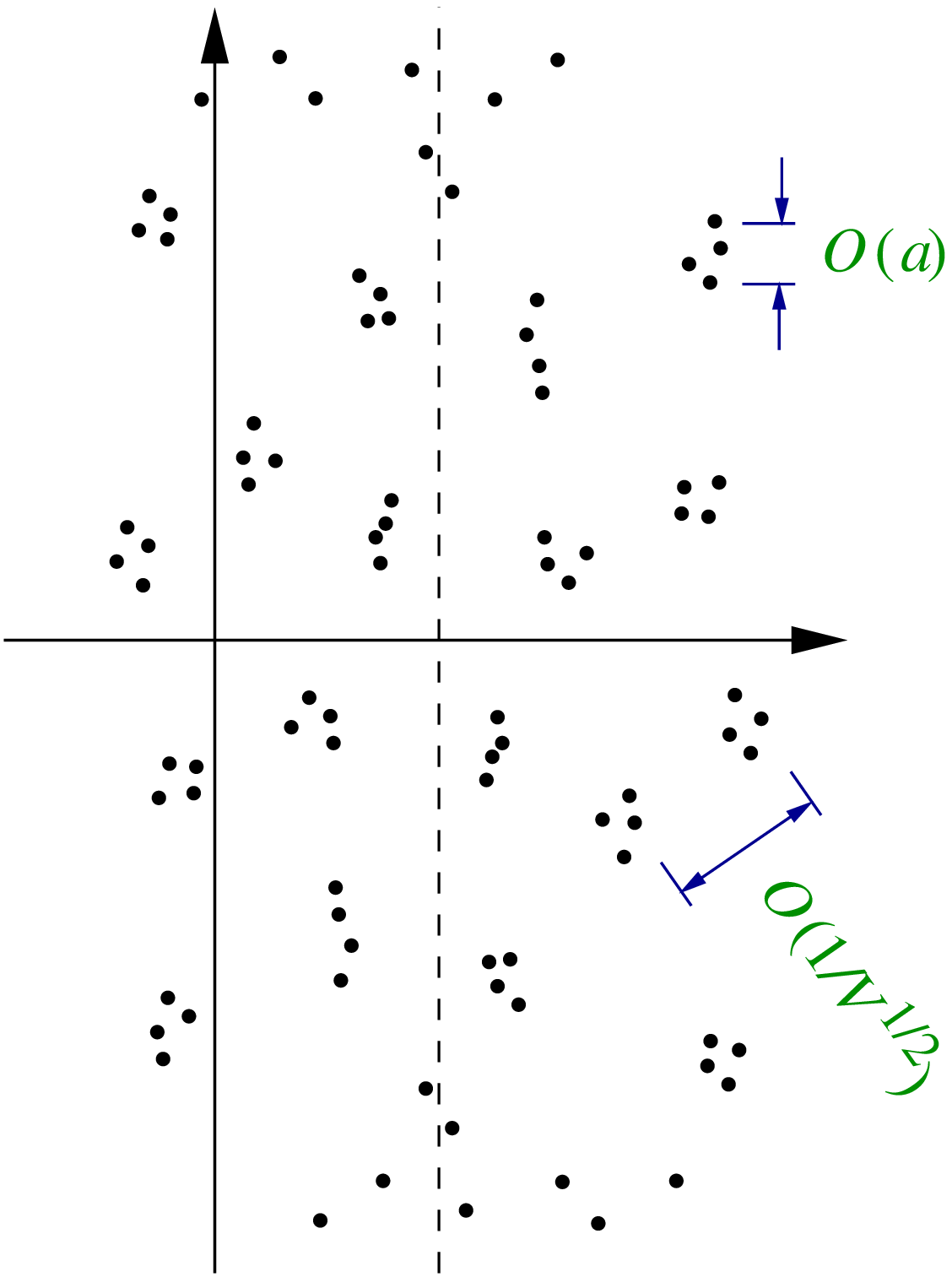}
\end{center}
\caption{Sketch of complex eigenvalues $\eta_i$ of $D+m$ when $\mu=0$ (left) and when $\Re\mu$
is somewhat larger than $m_\pi/2$ (right). The dashed line is $\Re\eta=m$.
\label{fig1}}
\end{figure}

When $\mu=0$ the Dirac operator $D[U]$ is anti-hermitian, so the eigenvalues of $D+m$ lie on a line parallel to the imaginary axis, $\Re\eta=m$ (see Fig.~\ref{fig1}).
If we are close to the continuum limit, the low-lying eigenvalues are near-degenerate because of the approximate taste symmetry; they are grouped into quartets corresponding to the four tastes, with an $O(a)$ splitting.\footnote{But see Sec.~5.}  When $\Re\mu>0$ the eigenvalues spread out into the complex plane as shown.
The density of eigenvalues (per unit area in the complex plane) is proportional to the 4-volume $V$; the splitting of each quartet is still $O(a)$.

How wide is the spread from the original $\Re\eta=m$ line?  The distribution can reach the imaginary axis for special configurations $U$ if $\Re\mu\gtrsim m$, the quark mass.%
\footnote{Our $\mu$ is the quark chemical potential, so the baryon chemical potential is $\mu_B=3\mu$.}
Such configurations, though, will be rare and atypical until $\mu$ is considerably larger.  Analysis in the $\epsilon$ regime \cite{Akemann} shows that the typical eigenvalue distribution will reach the axis (and the origin) when $\mu\gtrsim m_\pi/2$.  This prediction is for $T=0$; for $T>0$ the bound is expected to move upward, as it is connected to the phase transition in the phase quenched theory \cite{Kim}.

\section{Fourth roots}

\begin{figure}[hbt]
\begin{center}
\includegraphics*[width=.4\columnwidth]{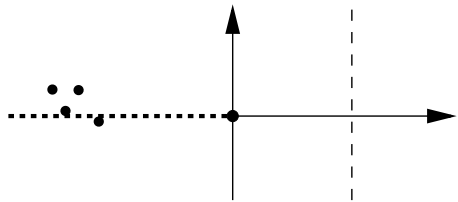}
\end{center}
\caption{An eigenvalue quartet as it crosses the cut, which was chosen to lie on the negative real axis.
\label{fig2}}
\end{figure}

Now let's see where phase jumps come from.  We can consider one quartet at a time, and define tentatively its contribution to $\Delta^{1/4}$ as a product of roots, each taken with a fixed cut:
\begin{equation}
\left(\prod_1^4\eta_i\right)^{1/4}=\prod_1^4\eta_i^{1/4}
\label{fixed}
\end{equation}
The phase of this product will jump by $\pi/2$ whenever any $\eta$ crosses the cut (Fig.~\ref{fig2}).

\begin{figure}[hbt]
\begin{center}
\includegraphics*[width=.4\columnwidth]{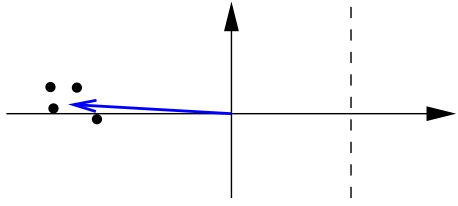}
\end{center}
\caption{Choosing the phase of the root in the ``ideal prescription.''
Now there is no need for a cut.
\label{fig3}}
\end{figure}

These jumps can be avoided by an alternative prescription.  Working with one quartet at a time, we define the phase of $(\prod\eta_i)^{1/4}$ to point to the center of the quartet, no matter where the quartet may be (Fig.\ref{fig3}).
Then the phase of the root follows the quartet around without jumps.  This is nothing other than the smooth replacement of four tastes by one quark per flavor---which is the whole idea of taking the fourth root!  We call this the ``ideal prescription.''

\begin{figure}[htb]
\begin{center}
\includegraphics*[width=.4\columnwidth]{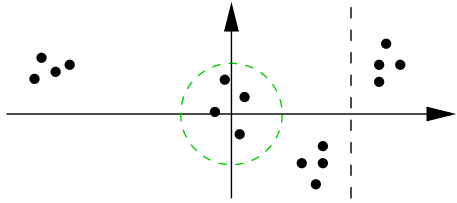}
\end{center}
\caption{An eigenvalue quartet that frames the origin.
\label{fig4}}
\end{figure}

The ideal prescription runs into trouble when applied to certain problematic configurations.  If there is a quartet that frames the origin (see Fig.~\ref{fig4}), there is no clear definition of the phase of the root.
Whether we keep such configurations or drop them, they will contribute a systematic error to any result.  The relative error will be the probability of finding such a quartet at the origin, which is the eigenvalue density times the area occupied by the quartet, or $O(a^2V\Lambda^6)$ for quenched configurations.  Reweighting reduces the contribution of these configurations by a factor of $O(a\sqrt{V}\Lambda^3)$, so the error estimate becomes $O[(a\sqrt{V}\Lambda^3)^3]$.

The devil is in the factors of volume in these error estimates.  They are connected with the fact that we must have $a\sqrt{V}\Lambda^3\ll1$ in order to have well-separated quartets.  The volume, however, cannot be reduced freely---it is usually fixed by physics requirements, such as $m_\pi L\gtrsim3$ for a thermodynamic ``limit.''  In order to control the error, it is essential to take the limit $a\to0$ before taking $V\to\infty$.

\section{Other prescriptions}

So much for our ideal prescription.  It requires identifying quartets and treating them individually.  If we don't do this, how bad can things get?

Let's go back to the fixed-cut prescription, \Eq{fixed}.  The phase of $\Delta^{1/4}$ jumps whenever an eigenvalue crosses the cut; it is restored to the correct value only when an entire quartet has crossed the cut.  Thus---if there are clean quartets in the first place---the systematic error comes from configurations where a quartet enters an exclusion zone of width $\sim a$ along the negative real axis.
The systematic error is thus $O(aV\Lambda^5)$. (For simplicity we take $\mu=O(\Lambda)$; in any case we take  $\mu\gtrsim m_\pi/2$, so that the eigenvalue density around the origin and to its left is appreciable.)

If we are not very close to the continuum limit, there will be no quartet structure apparent in the eigenvalue distribution.  The figure of merit is, again, $a\sqrt{V}\Lambda^3$.  If there are no quartets then the phase flips will be uncorrelated and the systematic error incalculable.

\section{Numbers}

Where are we today?  Let's begin with the figure of merit, $F\equiv a\sqrt{V}\Lambda^3$.  This is the ratio of the splitting in a quartet to the distance between quartets.  Let's require that it be less than, say, $1/5$.  If we fix the 4-volume $V=L^3/T$ by setting $L\gtrsim3m_\pi^{-1}$ and $T\sim T_c$, and identify the scale as $\Lambda\equiv T_c\simeq200$~MeV, we arrive at
\begin{equation}
a^{-1}>9\ {\rm GeV},
\label{cutoff}
\end{equation}
which should give one pause.  This is an order-of-magnitude estimate, which allows either optimism or pessimism.  The true test in any calculation will be to display the eigenvalue distribution of a typical configuration in order to see whether the quartet structure is evident.

We can judge the state of the art by the largest reweighting calculation done to date \cite{FK3}, wherein
$N_t=4$ and $T\simeq T_c$, so that $a^{-1}\simeq 800$~MeV, and $V\Lambda^4=27$.  This gives $F\simeq1.3$, meaning there can be no well-defined quartets to work with.
Moreover, the method of Ref.~\cite{FK3} is to write $\Delta=\prod\xi_i$ where the factors $\xi_i$ are {\em not\/} eigenvalues of $D+m$.  We know of no analysis that predicts a taste structure for the $\xi_i$ and thus no reason why quartets should form in the continuum limit.

Going a step further, if some calculation does manage to reach $a^{-1}\simeq9$~GeV (as it were), then $F^3$ gives an estimate of the systematic error of $O(1\%)$ for the ``ideal prescription.''
Recall that the ideal prescription requires identifying all quartets and taking their fourth roots carefully.
If, instead, one were to use the fixed-cut prescription in this regime in order to avoid identifying quartets, then the relative systematic error, calculated as $O(aV\Lambda^5)$, would still be $\simeq1.8$.  Bringing this down to, say, 10\% by decreasing the lattice spacing is a daunting prospect.

\section{\label{sec:aa2}$a$ or $a^2$?}

I have assumed that taste breaking in the eigenvalue spectrum is an $O(a)$ effect, which is the weakest assumption consistent with a continuum limit.
In discussions following this conference, the possibility arose that the splitting of the eigenvalue quartets might be $O(a^2)$ rather than $O(a)$.  I add a brief (and inconclusive) discussion of this issue here for completeness.  If indeed the splitting is $O(a^2)$ then our estimates of systematic errors change in the obvious way; I shall display the recalculated estimates below.

Let us begin with {\em free\/} staggered fermions.  At $\mu=0$ the fermion action may be rewritten in terms of block fields, rearranged in the ``taste basis,'' as \cite{taste}
\begin{equation}
S_{\rm taste}=(2a)^4\sum_{X}\overline \psi(X)\;
[\nabla_\mu(\gamma_\mu\!\otimes\!I)-a\Delta_\mu(\gamma_5\!\otimes\!\tau_\mu\tau_5)+m]
\;\psi(X).
\label{taste}
\end{equation}
Here $X$ labels a block of $2^d$ sites and the fermion field $\psi$ has 16 components, which are decomposed into a Dirac index (for $\gamma$ matrices) and a taste index (for $\tau$ matrices), each running from~1 to~4.
The first term gives a dispersion relation of naive fermions, replicated for the four tastes.  The second term, with an explicit $a$ in the coefficient, removes the doublers and splits the taste degeneracy.  Taking, for instance, $p_\mu=(p,0,0,0)$, with $pa\ll1$, we find that the first term takes the form $ip\gamma_1\otimes I$ while the second is $ap^2\gamma_5\otimes\tau_1\tau_5$.  The two terms don't commute.  Diagonalizing the operator then gives splittings that are of $O(p^2a^2)$ relative to the unperturbed energy $p$.  This result, which generalizes easily to any $p$, shows that the free theory's quartets are split in $O(a^2)$, and not as assumed above.

In the interacting theory, the gauge field insinuates itself into taste-invariant and taste-breaking terms alike, and it looks like there is no reason why splittings shouldn't be of the naive $O(a)$ according to the explicit $a$ in the second term in \Eq{taste}.  
(As noted in Ref.~\cite{GSS}, this is not inconsistent
with the well-known $a^2$ scaling of all physical observables.)  It is known, however~\cite{Luo}, that all $O(a)$ terms in
the interacting
action can be eliminated by switching to improved fields defined by
\begin{equation}
  \psi \to (1+{\cal T}) \psi ,
\qquad\qquad
  \overline \psi \to \overline\psi (1-{\cal T}) ,
\end{equation}
where ${\cal T}={\cal T}(U)$ connects nearest neighbors and is anti-hermitian.
This can be readily generalized to $\mu\ne 0$ (where ${\cal T}$ is
no longer anti-hermitian).  This follows~\cite{Luo,SharpeTASI} from the 
fact that the Symanzik effective action for staggered fermions equals the continuum
action plus terms of order $a^2$.   What is unclear however, is whether this also implies
that eigenvalue splittings are of order $a^2$, at least for eigenvalues of order $\Lambda$
or below (which are the ones at issue here \cite{GSS}).
The problem is that the usual Symanzik analysis of irrelevant operators applies
to the calculation of low-energy physical observables, {\em not\/}
to the calculation of individual eigenvalues of the Dirac operator.
It thus remains an open question whether taste splittings in the eigenvalues scale as $a$ or as $a^2$. 

If we suppose that the taste splitting is of $O(a^2)$, then the figure of merit becomes $F'\equiv a^2\sqrt{V}\Lambda^4=(a\Lambda) F$.  Requiring $F'<1/5$ gives [cf.~\Eq{cutoff}]
\begin{equation}
a^{-1}>1.3\ {\rm GeV},
\label{cutoff2}
\end{equation}
which is no longer astronomical.  A reassessment of the parameters used in Ref.~\cite{FK3}, moreover, brings its figure of merit down to  $F' \simeq 0.3$, which may offer a glimmer of hope.

Two things, however, are unchanged.  First, the relative error in the ideal prescription, $F^{\prime3}\simeq1\%$, applies only if one identifies quartets and treats them as indicated (the Fodor--Katz method, as we have mentioned, doesn't do this).  Second, if one accepts $F'=1/5$ but uses a fixed-cut prescription, then the relative error here ($a^2V\Lambda^6$) still turns out to be $\simeq1.8$, because $a$ is larger than before.  Even if one starts with $a^{-1}=1.3$\ GeV, this will have to be raised by a factor of $\sqrt{18}$ to reach an error of 10\% (the error, remember, is now $\sim a^2$).

Whichever error estimate we adopt, it can be reduced by using an improved action.  A tree-level improvement to the staggered action will reduce taste breaking by a factor of $\alpha_s$, and our estimates will change accordingly.  We don't see how this will change any orders of magnitude, but some improvement can be expected.

\section{Conclusions}

I will end by emphasizing three points.  First, the only way to see whether a calculation will have its fourth-root error under control is to see
whether the eigenvalues calculated for generated configurations really fall into well-defined quartets in a sizable region that includes the origin and the negative real axis.  (More distant eigenvalues can be treated by a fixed-cut prescription.)  This will render moot any arguments about our estimated figure of merit and about $a$ vs.~$a^2$.

Likewise, the whole problem arises only if $\Re\mu$ is sufficiently large that eigenvalues have begun to circle the origin in the complex plane.  I have quoted a rough bound $\Re\mu\gtrsim m_\pi/2$ for $T=0$
\cite{Akemann}, and noted that this bound should move upwards as $T$ is increased \cite{Kim}. Any calculation that claims to lie {\em below\/} the bound should prove it by, again, plotting eigenvalue distributions.

Finally, if the quartet structure is verified, one {\em cannot\/} just go ahead and use a fixed-cut prescription for all eigenvalues.  The error will still turn out large, because there will always be a quartet that straddles the cut and gives the wrong phase to the determinant.  One {\em must\/} use the ideal prescription, proceeding quartet-by-quartet to assign an unambiguous phase to the root of the product.

\acknowledgments
We thank Claude Bernard for discussion of the points covered in Section \ref{sec:aa2}, and also Zoltan Fodor for much discussion.
MG and BS thank the Benasque Center for Science for its hospitality.
This work was supported by the Israel Science Foundation under grant
no.~173/05 and by the US Department of Energy.

\end{document}